\documentclass[%
 superscriptaddress,
 nobibnotes,
 amsmath,amssymb,
 aps,
 reprint,
 physrev,
 longbibliography,
 floatfix
]{revtex4-2}

\usepackage{graphicx}
\usepackage{dcolumn}
\usepackage{bm}
\usepackage[caption=false]{subfig}

\usepackage{floatrow}
\begin{document}


\title{Compressibility measurement of the thermal MI--BG transition in an optical lattice}

\author{Phil Russ}
\altaffiliation[Present address: ]{Upgrade, San Francisco, California 94111, USA.}
\affiliation{Department of Physics, University of Illinois Urbana-Champaign, Urbana, Illinois 61801, USA}

\author{Mi Yan}%
\altaffiliation[Present address: ]{Wolfram Research, Champaign, Illinois 61820, USA.}
\affiliation{Department of Physics, Virginia Tech, Blacksburg, Virginia 24061, USA}

\author{Nicholas Kowalski}
\affiliation{Department of Physics, University of Illinois Urbana-Champaign, Urbana, Illinois 61801, USA}

\author{Laura Wadleigh}
\altaffiliation[Present address: ]{Atom Computing, Boulder, Colorado 80301, USA.}
\affiliation{Department of Physics, University of Illinois Urbana-Champaign, Urbana, Illinois 61801, USA}

\author{Vito W. Scarola}
\affiliation{Department of Physics, Virginia Tech, Blacksburg, Virginia 24061, USA}

\author{Brian DeMarco}
\email[Corresponding author: ]{bdemarco@illinois.edu}
\affiliation{Department of Physics, University of Illinois Urbana-Champaign, Urbana, Illinois 61801, USA}

\date{\today}

\begin{abstract}
Disorder can be applied to transform conducting to insulating states by localizing individual quantum particles.  The interplay between disorder and interactions in many-particle systems leads to a richer tapestry of quantum phase transitions. Here, we report the measurement in an ultracold lattice gas of a disorder-induced transition from a state with small disorder-independent compressibility to a state for which compressibility increases with disorder. At zero temperature this is the transition from a Mott insulator (MI) to a Bose glass (BG), both of which are insulating states. This transformation is observed using measurements of core compressibility. By determining how double occupancy changes with atom number, we identify the threshold disorder strength required to switch from disorder-independent MI-like to disorder-dependent BG-like compressible behavior. 
\end{abstract}

\maketitle

\section{Introduction}

The interplay between disorder and interactions leads to a rich variety of many body physics and is essential to understanding the behavior of many condensed matter systems. The disordered Bose-Hubbard model (DBHM) is a paradigm for understanding this interplay. Here, we present a measurement of a disorder induced transition from a disorder-independent low compressibility state to a state with compressibility that increases with disorder. This transition is the finite temperature remnant of the Mott insulator (MI)--Bose glass (BG) transition. Both phases are insulating. However, the zero-temperature MI is incompressible with a gapped excitation spectrum, while the zero-temperature BG is compressible and gapless. The compressibility behavior we present here is a key characteristic of this transition. 

Atomic lattice gas experiments have proven to be tremendously powerful tools for studying strongly correlated physics and, including the Bose-Hubbard model. An early landmark in the success of this type of simulation was the observation of the superfluid (SF)--MI transition \cite{Greiner2002,Stoferle2004,Spielman2007}. The addition of disorder enabled observation of the SF--BG transition in one- \cite{Tanzi2013,DErrico2014a} and three-dimensions \cite{Pasienski2010,Meldgin2016}. More recent results include a measurement of the SF--BG transition in a two-dimensional quasicrystal \cite{Yu2023} and the disorder-induced delocalization of a one-dimensional MI \cite{Yao2024}. 

 We focus on the MI and BG phases of the DBHM in the regime of strong interactions ($U$) and  large disorder ($\Delta$). While both phases are insulators, the finite excitation gap of the MI means it is incompressible, whereas the BG is gapless and thus compressible at $T=0$. The MI--BG phase transition is of the Griffiths type: particles are able to delocalize within exponentially rare regions where the disorder potential sufficiently offsets the interaction-induced excitation gap \cite{Gurarie}. While exceedingly rare, these regions nonetheless have finite probability of occurring in an infinite system. The reliance of this scenario on large systems sizes implies that this phase transition may be inaccessible to exact numerical methods and raises questions about what measurements would reveal for physical systems. An example of the ground state phase diagram at temperature $T=0$ for the 3D uniform system at unit filling is computed using quantum Monte Carlo (QMC) in Ref. \cite{Gurarie} for box disorder. In this scenario the MI--BG transition is inaccessible to calculation but is conjectured to occur at a disorder strength of $\Delta=U/2$ \cite{Pollet2009}.

Measurements probing the MI--BG transition have been performed in one dimension \cite{Gadway2011,DErrico2014b}. Transport measurements confirming an insulating state and modulation spectroscopy of the closing excitation gap hint at the transition to a BG state with the addition of disorder. However, the compressibility was not measured in these previous works, a key feature of this transition. Another difference is the type of disorder used, with these works using incommensurate lattices and impurity atoms to realize disorder, not the optical speckle pattern we present here. 

In this work, we probe the core compressibility of strongly interacting bosonic atoms trapped in a cubic disordered lattice. We measure the fraction of atoms on doubly-occupied sites $D$ in a small window of total atom number $N$ around unit filling at the center of the gas in order to infer the core compressibility ratio $\kappa_c/\kappa=\partial D/\partial N$ \cite{Scarola,Khorramzadeh}, where $\kappa_c$ and $\kappa$ are the core and total compressibilities, respectively. In the presence of an overall harmonic trapping potential, this quantity allows us to isolate the behavior of the center of the gas where the physics resembles that of the uniform system. Without this method, the MI--BG transition would be obscured by contributions from the edge of the gas where the compressibility is always finite. Using $D$ to probe phase transitions in Hubbard models has been established experimentally for both bosons \cite{Gerbier} and fermions \cite{Jordens}.

For fixed values of the interaction strength, we observe the behavior of $\kappa_c/\kappa$ over a range of finite disorder strengths and compare to theoretical predictions. We observe a disorder-induced finite temperature transition from a disorder-independent state to one where compressibility increases with increasing disorder. Comparison to atomic-limit predictions confirm that this is consistent with the finite-temperature remnant of the MI--BG quantum phase transition (QPT). To confirm the validity of the atomic limit approximation in this regime, Gutzwiller variational wavefunction (i.e., site-decoupled mean-field theory) and QMC calculations were also performed (see Supplemental Material for more information \cite{supp}). 

\section{Measuring Double-Occupancy}

We prepare a BEC of $^{87}$Rb atoms, which are trapped in a cubic disordered optical lattice formed by three pairs of retro-reflected $\lambda_L=812$ nm laser beams and an optical speckle field created using a $\lambda_D=532$ nm laser beam (see the methods section for details of the experimental sequence). This system, under certain approximations, is described by the disordered Bose--Hubbard Hamiltonian \cite{Fisher}
\begin{eqnarray} 
\mathrm{H}=-\sum_{\langle ij\rangle}t_{ij} (\hat{b}^\dagger_i \hat{b}_j + \hat{b}^\dagger_j \hat{b}_i)  + \frac{1}{2} \sum_i U_i \hat{n}_i (\hat{n}_i-1)  \nonumber\\ +\sum_i (\epsilon_i + \Omega^2 r_i^2) \hat{n}_i,
\label{eq:BoseHubbard}
\end{eqnarray}
where sites are labeled $i$ and $j$, and $\hat{b}^\dagger_i$ ($\hat{b}_i$) and $\hat{n}_i = \hat{b}^\dagger_i \hat{b}_i$ are the boson creation (annihilation) and site occupancy operators, respectively. Tunneling, with energy $t_{ij}$, occurs between nearest-neighbor sites as denoted by the sum index $\langle ij\rangle$ and two-body interactions, with strength $U_i$, occur between atoms on the same site. The last term in Eq. \ref{eq:BoseHubbard} describes the site occupation energy shift stemming from two sources. The disorder potential causes random shifts $\epsilon_i$ and is characterized by the disorder strength $\Delta$, which is approximately both the mean and standard deviation of the $\epsilon_i$. Also, the overall harmonic confinement shifts the energy in proportion to $\Omega^2=m\omega^2/2$, where $\omega$ is the trap frequency for an particle of mass $m$ at a distance $r_i$ from the trap center. We quantify the Hubbard parameters with respect to the lattice recoil energy $E_R=h^2/\left(2m\lambda_L^2\right)\approx 170~\mathrm{nK} \times k_B$.

\begin{figure}[htp]
\includegraphics[width=0.95\linewidth]{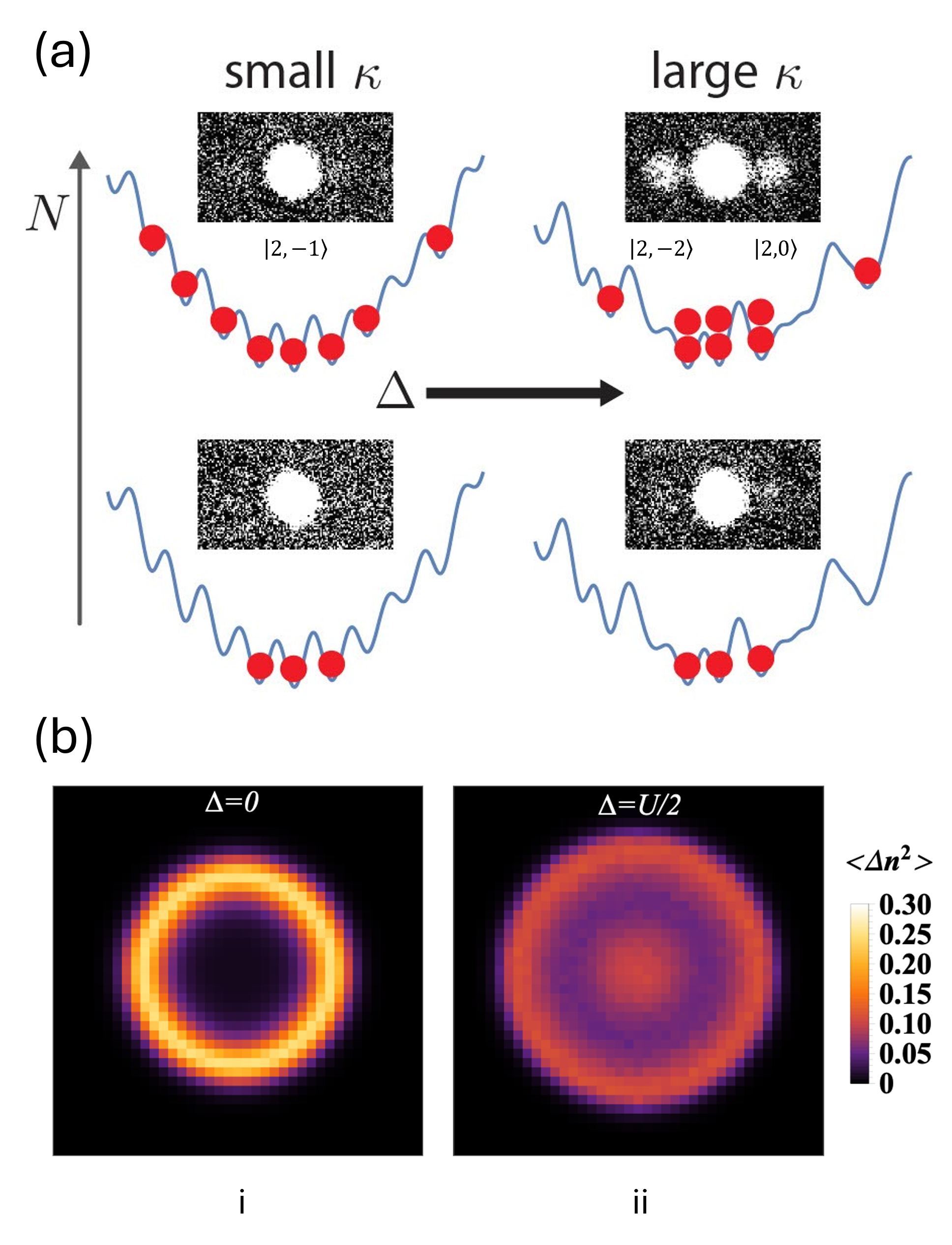}
\caption{Overview of the compressibility measurement. (a) Time of flight images taken at $s=20$~$E_R$. The top row shows atom number $8700$, the bottom $N=5400$. The left column shows $\Delta=0$, and the right $\Delta=0.3$~$E_R$. Spin exchange interactions result in atoms on doubly-occupied sites transferring from the $\big|2,-1\rangle\otimes\big|2,-1\rangle$ hyperfine state to the $\big|2,-2\rangle\otimes\big|2,0\rangle$ hyperfine state. (b) Plot of the mean-field theory predicted density fluctuations in the $x-y$ plane of a 2D trapped lattice, obtained by using the Gutzwiller approximation with $U=0.52 E_R$, $\Omega^2d^2/U=4.7\times10^{-3}$ (where $d=406$~nm is the lattice site spacing), and $T=20t=0.096U$. The results were disorder averaged over 1000 realizations.  Panel i, at $\Delta=0$, shows essentially no density fluctuations near the trap center, consistent with the Mott regime.  Panel ii adds enough disorder, $\Delta=U/2$, to drive the center of the system into the compressible regime. The total compressibility is large and nearly the same for both panels because the system edges remain compressible.}
\label{fig1}
\end{figure}


In this experiment, we achieve entropies such that $t \ll k_b T \ll U$, where the behavior of the compressibility can be understood solely in terms of the competition between interactions and disorder-induced thermal fluctuations due to disruption of the MI gap. In this regime, the gap suppresses thermal fluctuations within the MI, causing entropy to concentrate at the edge. The MI compressibility is activated exponentially with an energy scale set by the gap $E_g \approx U$. We constrain the entropy-per-particle $S/N$ of the system by measuring $D$ over a range of $N$ at $\Delta=0$ and comparing to predictions using the atomic limit. For our system, this results in a nearly incompressible thermal MI with $S_{core}/N < S_c/N$, where $S_{core}/N$ is the entropy-per-particle in the core, and $S_c/N$ is the BG critical entropy-per-particle for the uniform system using a Gutzwiller mean-field description \cite{supp}.

The addition of disorder disrupts the MI by introducing states into the gap, allowing entropy to redistribute from the edge to the core. This gradually increases thermal fluctuations which control double occupancies and the core compressibility as illustrated in Fig. \ref{fig1}. For small $\Delta$, thermal fluctuations remain small causing the core compressibility and double occupancies to remain suppressed. For large $\Delta$, large thermal fluctuations activate the core compressibility and causes double occupancies to proliferate. This is also qualitatively consistent with a simple physical picture where double occupancies become energetically favorable as $\Delta \to U$. In this experiment, $S_{core}/N > S_c/N$ for finite $\Delta$ due to adiabatic heating in the core and instead of a QPT, there is a crossover from the thermal MI to the thermal BG.

To measure double occupancy in the system---and thus compressibility---we use spin exchange in the $F=2$ hyperfine ground state. A BEC is initially prepared in the $\big|1,-1\rangle$ hyperfine state and loaded into the disordered lattice potential. Before measuring $D$, the lattice depth is ramped to 40~$E_R$ in 0.5~ms to halt all dynamics. A low magnetic field breaks the degeneracy of the magnetic Zeeman states and acts as a switch for the spin exchange process. The atoms are transferred to the $|2,-1\rangle$ hyperfine ground state via adiabatic rapid passage using an applied microwave-frequency magnetic field. After state preparation, the low magnetic field is ramped off, allowing spin exchange to begin. On doubly occupied sites, the initial state is $|2,-1\rangle \otimes |2,-1\rangle$, leaving $|2,-2\rangle \otimes |2,0\rangle$ as the only accessible final state due to conservation of total magnetization. After waiting for one $\pi$-time of the spin exchange oscillation, a vertical magnetic field gradient is snapped on and bandmapping \cite{McKay2009} is performed by ramping off the lattice potential in 0.2~ms. This magnetic field gradient serves the dual purpose of halting the spin exchange process and to spatially separate the spin states during time of flight, as shown in figure \ref{fig1}. 

\begin{figure}[htp]
    \centering
    \includegraphics[width=0.9\textwidth]{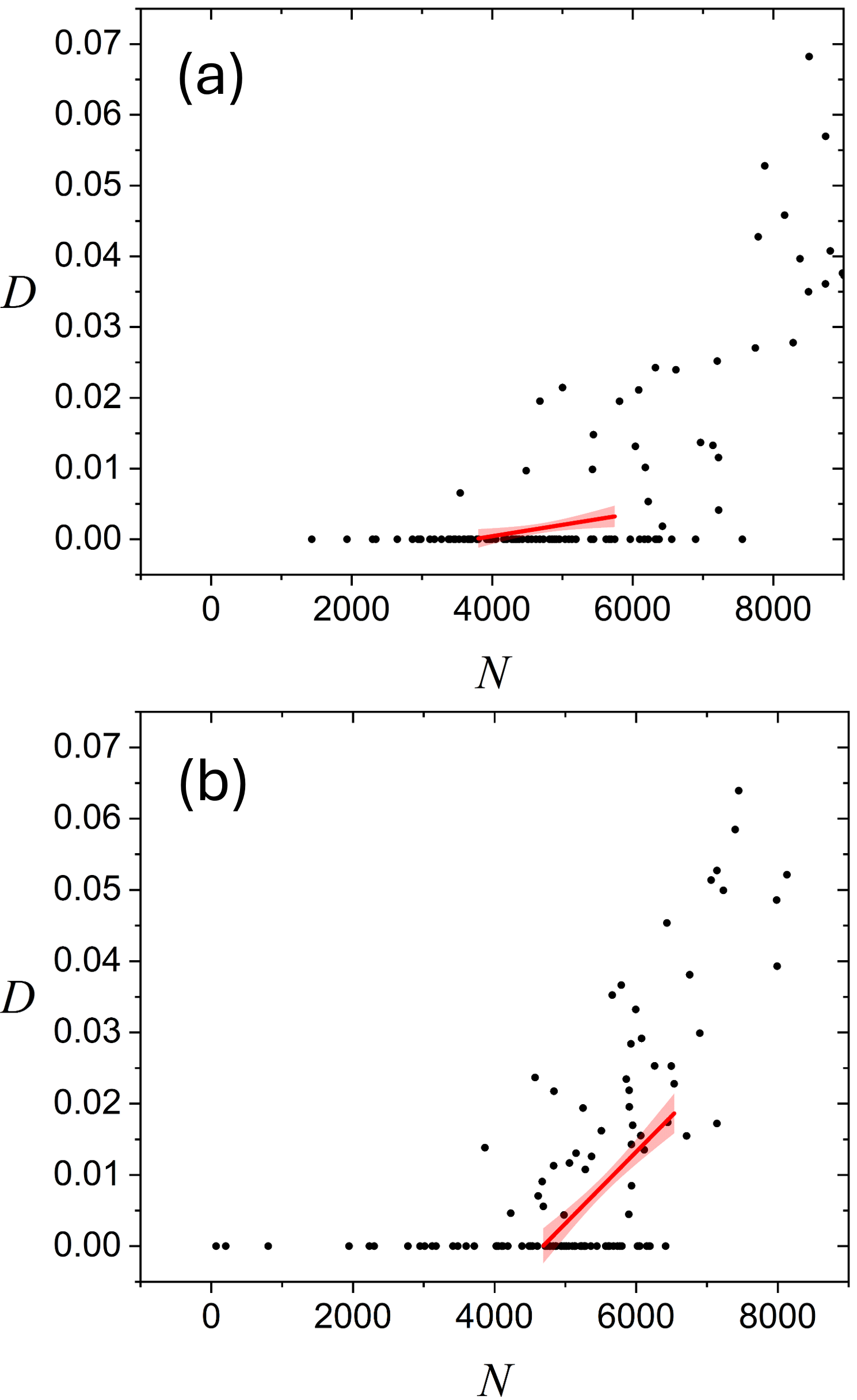}
    \caption{Measured $D$ at $s/E_R=20$ for $S/N=0.84\ k_B$ and $\omega/(2\pi)=90$ Hz. (a) $\Delta/E_R=0$ and (b) $\Delta/E_R=0.16$. The points at $D=0$ are the result of a machine learning classifier indicating the absence of atomic clouds in the regions where double occupancies show up in TOF images. The red line is a linear fit to the data in a window of $|N-N_1|\leq1000$ and the red shaded region is the 68\% confidence interval of the fit. The slope of the linear fit is used as the measured core compressibility ratio $\kappa_c/\kappa$.}
    \label{fig2}
\end{figure}

In Fig. \ref{fig2} we show two examples of measurements for $D$ at $U/E_R=0.52$. We quantify double occupancies by analyzing time-of-flight absorption images with a two-step method which identifies images where $D=0$ and otherwise determines $D$ and $N$ from pixel summing. From this analysis, we obtain $N=\sum_{m=-2}^{0} N_{|2,m\rangle}$ and $D=(N_{|2,-2\rangle} + N_{|2,0\rangle})/N$, where $N_{|2,m\rangle}$ is the population in the $|2,m\rangle$ ground hyperfine spin state. More information on the analysis technique is provided in the methods section. Focusing on $D$ near the atom number corresponding to unit fulling $N_1=N(\mu=0.5U+\Delta)$ (for speckle disorder) where $\mu$ is the global chemical potential, we observe similar behavior for all $U$. At small $\Delta$, $D$ is suppressed for s/mall changes $N$, as expected for the persistence of a thermal MI. For large $\Delta$, we observe increasing $D$, which is consistent with the presence of large thermal fluctuations due to the disruption of the gap. Shown in Fig. \ref{fig1} are plots of the density fluctuations in a 2D plane using Gutzwiller mean-field theory. They show essentially no density fluctuations for $\Delta=0$, while for large $\Delta$, density fluctuations are present because the core has become compressible.

\section{Determining Compressibility}
\begin{figure}[htp]
\centering
\includegraphics[width=0.95\linewidth]{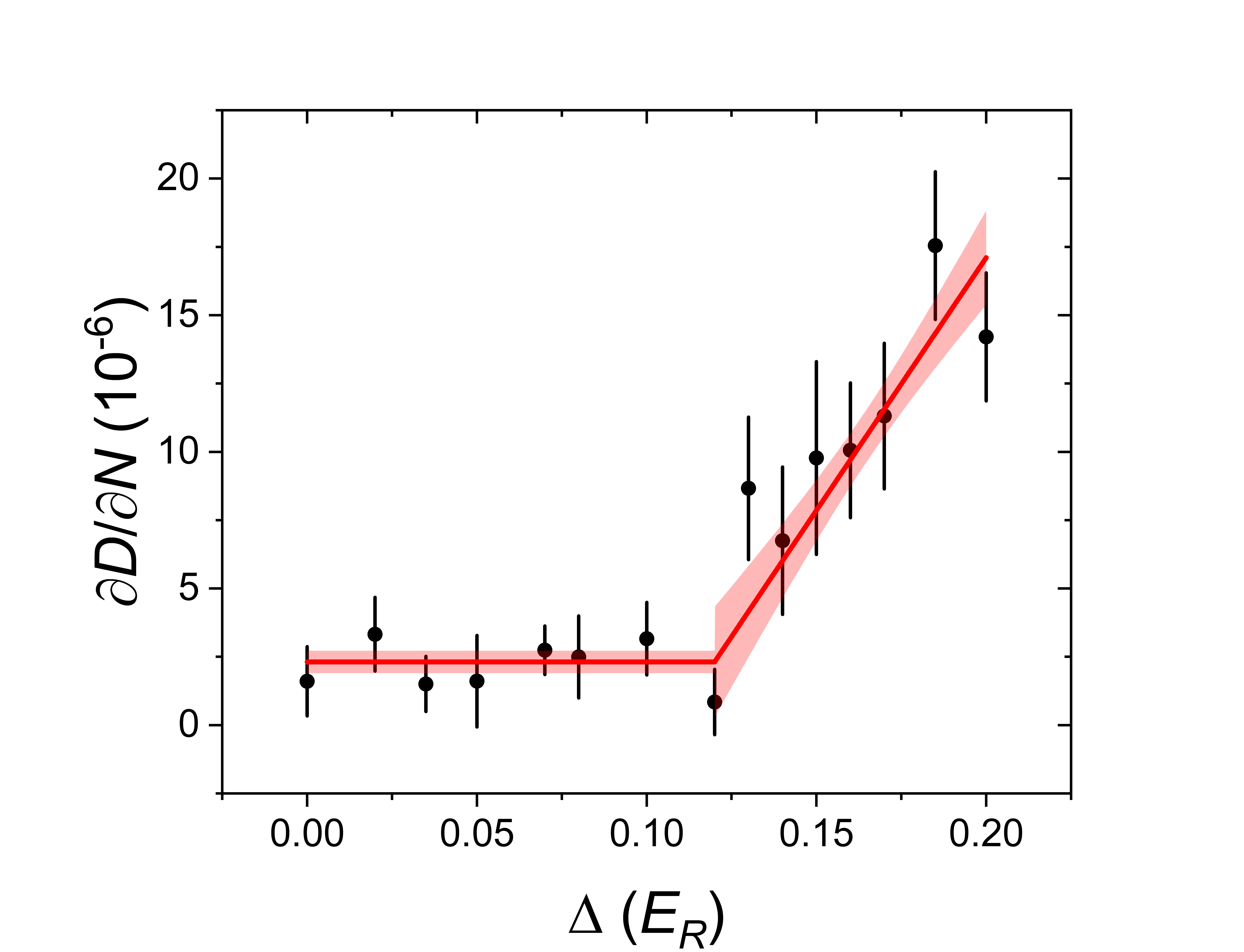}
\caption{Measured $\partial D/\partial N$ at $s/E_R=20$. The red line is a piecewise linear fit to the data and the red shaded region is the 68\% confidence interval of the fit. The discontinuity indicates when $\partial D/\partial N$ rises above the baseline value at small $\Delta$.}
\label{fig:Compressibility}
\end{figure}

We extract $\partial D/\partial N$ from the measurement at each $\Delta$ for fixed $U$ using a heuristic method where the measured $D$ is selected within a window of $|N-N_1|\leq1000$, and we obtain the slope from a linear least-squares fit to the data.
Fig. \ref{fig:Compressibility} shows the result of applying this analysis to the measured $D$ for $U/E_R=0.52$. For small $\Delta$, we observe a region of small $\partial D/\partial N$ that is independent of $\Delta$, within the uncertainty in the data. For large $\Delta$, we observe steadily increasing $\partial D/\partial N$. This is trend is reminiscent of the MI--BG phase transition, but with the partially compromised gap of the thermal MI remaining resilient over a small range of $\Delta$ before the thermal BG takes over. We observe a factor of 5--10 increase in $\partial D/\partial N$ from small to large $\Delta$ for all $U$. We analyze the measured $\partial D/\partial N$ using piecewise linear least-squares fitting. While a piecewise function is appropriate for low-temperature physics in the large-system limit, the size of the scatter and uncertainty in the data prevents differentiating between continuous and discontinuous behavior.

The threshold disorder, $\Delta_{th}$, quantifies the onset of increasing $\partial D/\partial N$ above the baseline value for small $\Delta$. We constrain the linear function below $\Delta_{th}$ to have zero slope to capture the near-constant behavior for small $\Delta$. 

We make a prediction of the threshold disorder as a function of $U$ in the atomic limit. 
To confirm the validity of the atomic limit, the results where compared to Gutzwiller variational wavefunction (site-decoupled mean-field theory) and QMC \cite{supp}. Using the partition function in the atomic limit, the relationship between $D$ and $N$ at the core was obtained by sampling a small region around the central site, using the same parameters as the experiment, including the harmonic trapping potential. The slope was obtained by averaging over 100 disorder realizations. 

\begin{figure}[htp]
\centering
\includegraphics[width=0.95\linewidth]{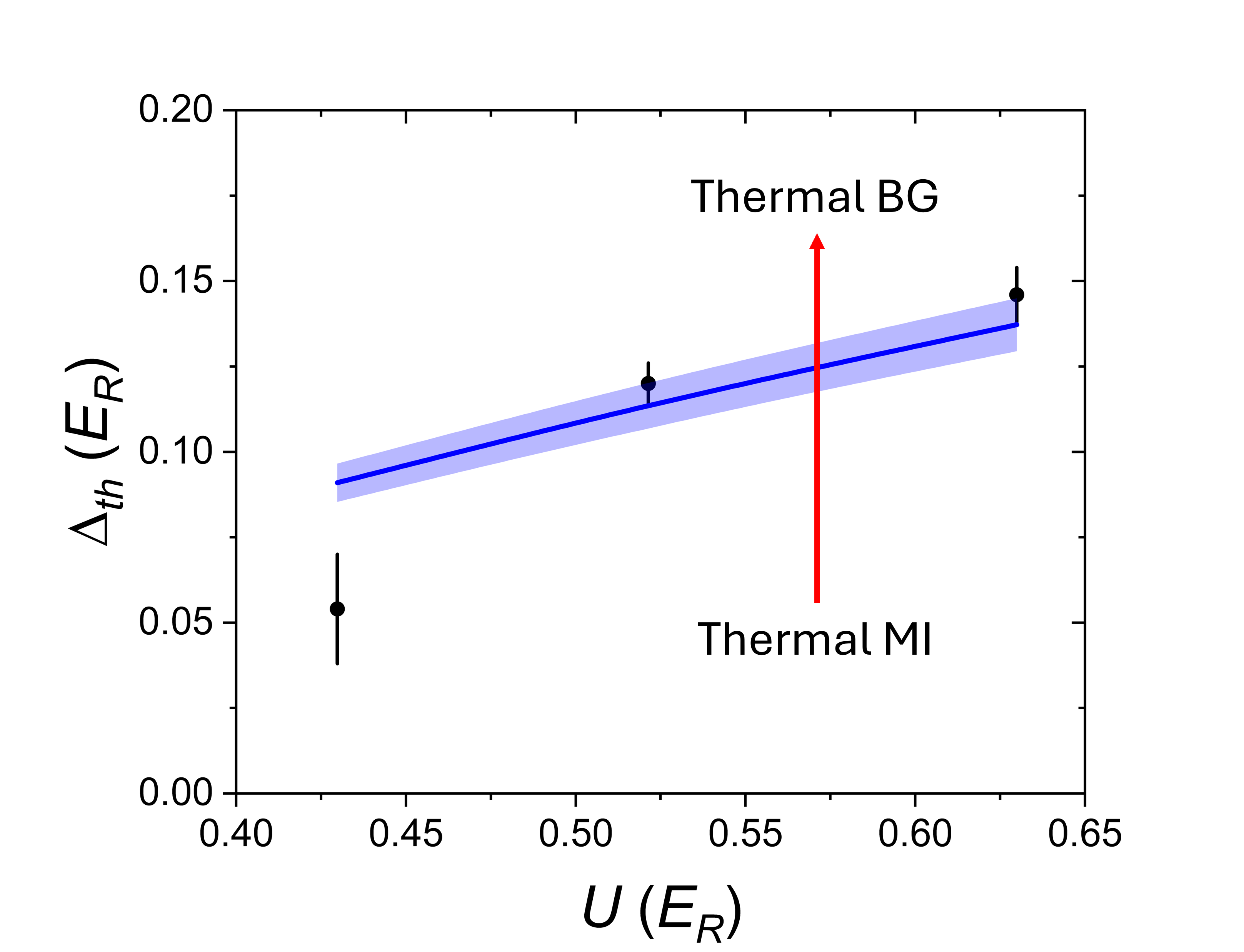}
\caption{Comparison of $\Delta_{th}$ for different $U$, where the measured value is the black points, and the theory prediction is the blue line. The systematic error in measuring $S/N$ in the experiment is the largest source of uncertainty and is shown as the blue shaded region in the predicted $\Delta_{th}$.}
\label{fig:ThresholdDisorder}
\end{figure}

To make a comparison between experiment and theory, we sample the predicted $\kappa_c/\kappa$ according to the uncertainty in the experimental data to generate simulated data and analyze it using the same piecewise linear least-squares fitting procedure to obtain a predicted $\Delta_{th}$. The comparison of $\Delta_{th}$ between experiment and theory is shown in Fig. \ref{fig:ThresholdDisorder}, where the theory prediction is computed for the experimental average $S/N=(0.75\pm0.03) \times k_B$. At lower lattice depths tunneling becomes a more significant factor, which is a likely explanation for the disagreement between the experimental data and the atomic limit prediction at the lowest value of $U$. We find good agreement with theory, indicating that the introduction of disorder disrupts the gap, driving the crossover from a thermal incompressible state to a thermal compressible state. 

\section{Conclusion}
We have achieved a measurement of an incompressible--compressible transition in a strongly correlated disordered many body system, using an atomic lattice gas to realize the disordered Bose-Hubbard model. This transition is a key feature of the predicted zero-temperature Mott insulator--Bose glass transition. Predictions using the Gutzwiller approximation find that the MI gives way to a quantum BG at the core when the critical entropy per particle is below a critical value of $(S/N)_{crit}=0.05~k_B$, which is significantly smaller compared to the entropies we realize here \cite{supp}. Further work repeating this measurement at lower entropy may realize the quantum MI--BG phase transition. 

\section{Acknowledgments}
This work is supported by the National Science Foundation through award 2110291.

P.R. and M.Y. contributed equally to this work. N.K. contributed to investigation and writing. 

\nocite{*} 
\bibliography{Compressibility}

\end{document}